\documentclass[12pt]{article}
\usepackage{amssymb}
\usepackage{amsmath}
\usepackage{amscd}
\usepackage{latexsym}
\def\a{\alpha}

\def\k{\kappa}

\def\be{\begin{equation}}
\def\ee{\end{equation}}
\def\arr{\begin{array}{rll}}
\def\ea{\end{array}}
\def\bea{\begin{eqnarray}}
\def\eea{\end{eqnarray}}
\def\N2{$N{=}2$}

\def\>{\rangle}
\def\<{\langle}
\def\+{\dagger}
\def\={\ =\ }

\begin{document}
%\large
\renewcommand{\thefootnote}{\fnsymbol{footnote}}

\begin{titlepage}
\setcounter{page}{0}
\begin{flushright}
LMP-TPU--14/13  \\
\end{flushright}
\vskip 1cm
\begin{center}
{\LARGE\bf  Reducibility of Killing tensors   }\\
\vskip 0.5cm
{\LARGE\bf in $d>4$ NHEK geometry }\\
\vskip 2cm
$
\textrm{\Large Dmitry Chernyavsky  \ }
$
\vskip 0.7cm
{\it
Laboratory of Mathematical Physics, Tomsk Polytechnic University, \\
634050 Tomsk, Lenin Ave. 30, Russian Federation} \\
{E-mail: recnart@list.ru}

\end{center}
\vskip 1cm
\begin{abstract} \noindent

An extremal rotating black hole in arbitrary dimension, along with time translations and rotations, possesses a number of hidden symmetries characterized by the second rank Killing tensors. As is known, in the near horizon limit the isometry group of the metric is enhanced to include the conformal factor $SO(2,1)$. It is demonstrated that for the near horizon extremal Kerr (NHEK) geometry in arbitrary dimension one of the Killing tensors decomposes into a quadratic combination of the Killing vectors corresponding to the conformal group, while the remaining ones are functionally independent.

\end{abstract}

\vspace{0.5cm}


\end{titlepage}

\renewcommand{\thefootnote}{\arabic{footnote}}
\setcounter{footnote}0

\noindent
{\bf 1. Introduction}

\vspace{0.3cm}

Killing tensors are conventionally attributed to hidden symmetries as there are no coordinate transformations in spacetime associated with them.
Along with Killing vectors, they provide important geometric characterization of a spacetime. As far as physical applications are concerned, they entail extra integrals of motion for the geodesic equations, which ensure separation of variables and integration of the equations in quadratures. A well--known example is the geodesic motion of a particle on the Kerr background \cite{carter,penrose}. In field theory, they are responsible
for the separation of variables in the Klein--Gordon and Dirac equations on the curved background \cite{carter2} (for a review see \cite{fk} and references therein).

A spacetime may also admit an antisymmetric analogue of the Killing tensor known in the literature as the Killing--Yano tensor.
In general, the Killing--Yano tensors may be used to construct the Killing tensors, but not every Killing tensor is decomposable into a combination of the Killing--Yano tensors (see e.g. \cite{frolov}). It is also worth mentioning that, when considering a superparticle model on a curved background which admits the Killing--Yano tensors,
extra supersymmetry charges can be constructed \cite{grh} which yield Killing tensors under the Poisson bracket.

Hidden symmetries of the Kerr--NUT--AdS black hole in arbitrary dimension and separation of variables in the geodesic equations were studied in a series of recent works \cite{separabil}--\cite{desert}. In particular, it was shown that in $d=2n+\epsilon$, where $\epsilon=1$ for an odd--dimensional spacetime and $\epsilon=0$ for an even--dimensional spacetime, there are $n$ functionally independent second rank Killing tensors, which also include the metric as a trivial Killing tensor. Along with time translations and rotations these guarantee the complete integrability of a particle model on such a background.

A salient feature of an extremal rotating black hole in arbitrary dimension is that one can consider the near horizon limit which yields a new vacuum solution of the Einstein equations. The latter differs from the parent geometry in many respects. The near horizon geometry is not asymptotically flat. The time translation symmetry is enhanced to the conformal group $SO(2,1)$  \cite{bardeen}. The latter paved the way to the Kerr/CFT--correspondence \cite{str} (for a review see \cite{com}) and to the extensive study of the conformal mechanics models associated with the near horizon black holes \cite{kallosh}--\cite{KO}. For the case of four dimensions it was also demonstrated \cite{g2,go} that the near horizon Killing tensor is reducible and can be expressed via a quadratic combination of the Killing vectors. Similar reducibility occurs for the Kerr--NUT--AdS black hole \cite{jr} and for the weakly charged extremal Kerr throat geometry \cite{fr}.

As was mentioned above, the number of second rank Killing tensors grows with dimension. It is then important to study how many of them are functionally independent in the near horizon limit. The purpose of this paper is to demonstrate that for the class of extremal rotating black holes described by the Myers-Perry solution \cite{mp} (vanishing cosmological constant and the NUT charge)
only one near horizon Killing tensor decomposes into a quadratic combination of the Killing vectors, while the remaining ones are functionally independent.

The paper is organized as follows. In Sect. 2 we first briefly review the construction of the second rank Killing tensors for an extremal rotating black hole in $d=2n$. The near horizon limit is discussed next and the reducibility of one of the Killing tensors is demonstrated. Similar analysis of the odd--dimensional case is carried out in Sect. 3.

\vspace{0.5cm}

\noindent
{\bf 2. Near-horizon Killing tensors in $d=2n$}

\vspace{0.3cm}

Our starting point is the Myers-Perry black hole solution \cite{mp} written in the coordinates introduced in \cite{general kerr}:
\bea\label{me2n}
&&
ds^2=\frac{U}{X} dr^2+\sum_{\a=1}^{n-1}\frac{U_\a}{X_\a} dx_\a ^2-\frac{X}{U} \Big[dt-\sum_{i=1}^{n-1}\frac{\gamma_i}{\varepsilon_i}d\phi_i\Big]^2+\sum_{\a=1}^{n-1} \frac{X_\a}{U_\a}\Big[dt-\sum_{i=1}^{n-1} \frac{(r^2+a_i^2)\gamma_i}{(a_i^2-x_\a^2)\varepsilon_i} d\phi_i\Big]^2,
\nonumber\\[2pt]
&&
U=\prod_{\a=1}^{n-1} (r^2+x_\a^2),   \qquad U_\a=-(r^2+x_\a^2) \prod_{\begin{subarray}{c} \beta=1\\\a\ne \beta\end{subarray} }^{n-1} (x_\beta^2-x_\a^2),
\nonumber\\[2pt]
&&
X_\a=-\prod_{k=1}^{n-1} (a_k^2-x_\a^2), \qquad  X=\prod_{k=1}^{n-1} (r^2+a_k^2)-2Mr ,
\nonumber\\[2pt]
&&
\gamma_i=\prod_{\a=1}^{n-1} (a_i^2-x_\a^2) \qquad \varepsilon_i=a_i\prod_{\begin{subarray}{c} k=1\\ k\ne i\end{subarray}}^{n-1} (a_i^2-a_k^2),
\eea
where $t$ is the temporal coordinate, $r$ is the radial coordinate and $x_\alpha$, $\phi_i$ with $i,\alpha=1,\dots,n-1$ are related to the angular variables.
In Eq. (\ref{me2n}) $M$ is the mass and $a_i$ are the rotation parameters.
Without loss of generality, it can be assumed that $a_1\leq a_2 \dots \leq a_{n-1}$ in which case the coordinates $x_\alpha$ take their values in the intervals   $a_\alpha\leq x_\alpha \leq a_{\alpha+1}$.

In what follows, we will also need the metric (\ref{me2n}) written in special coordinates introduced in Ref. \cite{general kerr}. It is obtained by the linear change of the variables
\bea
&&
B_{j}^{(k)}\equiv\sum_{\begin{subarray}{c} l_1<l_2\dots<l_k \\j\ne {l_1, \dots, l_k}\end{subarray} } a_{l_1}^2 a_{l_2}^2 \dots a_{l_k}^2, \qquad \frac{\phi_j}{a_j}=\sum_{k=0}^{n-2} B{_j}^{(k)} \psi_{k+1},
\nonumber\\[2pt]
&&
B^{(k)}\equiv\sum_{l_1<l_2\dots<l_k} a_{l_1}^2 a_{l_2}^2 \dots a_{l_k}^2, \qquad t=\psi_0 +\sum_{k=1}^{n-1}B^{(k)} \psi_k,
\eea
which yields \cite{general kerr}
\be\label{metr}
\ ds^2=\sum_{\mu=1}^{n} \ \bigg\{ \frac{dx_\mu}{Q_\mu}+Q_\mu \Big(\sum_{k=0}^{n-1} A_{\mu}^{(k)} d\psi_k \Big)^2 \bigg\},
\ee
where
\bea\label{funct 2n}
&&
Q_\mu=\frac{X_\mu}{U_\mu}, \qquad A_{\mu}^{(k)}=\sum_{\begin{subarray}{c} \nu_1<\nu_2\dots<\nu_k \\\mu\ne {\nu_1,\dots,\nu_k}\end{subarray} }^{n} x_{\nu_1}^2 x_{\nu_2}^2 \dots x_{\nu_k}^2,
\nonumber
\eea
\bea
&&
U_\mu=\prod_{\begin{subarray}{c} \nu=1\\\mu\ne \nu\end{subarray} }^{n} (x_\nu^2-x_\mu^2), \qquad X_\mu =-\prod_{k=1}^{n-1} (a_k^2-x_\mu^2)-2 \widehat{M} x_\mu \delta_{\mu n},
\nonumber\\[2pt]
&&
x_n=\texttt{i}r, \qquad \qquad\qquad ~ \widehat{M}=\texttt{i}M,
\eea
and $A_{\mu}^{(0)}\equiv1$.
In this coordinate system, the full set of $n$ functionally independent Killing tensors was constructed in Ref. \cite{desert}
\be\label{killing}
\emph{K}_{(k)}=\sum_{\mu=1}^{n}\Big[\frac{A_{\mu}^{(k)}}{X_\mu U_\mu} \Big(\sum_{j=0}^{n-1} (-x^2_{\mu})^{n-1-j} \partial_{\psi_j} \Big)^2+A_{\mu}^{(k)} Q_\mu(\partial_{x_\mu})^2\Big],
\ee
where $k=0, \dots, n-1$ and $K_{(0)}$ stands for the metric inverse to (\ref{metr}).

As the next step, let us discuss the near horizon limit of the metric (\ref{me2n}). In order to describe it, one considers the extremal case, for which $X$ has a double zero at the horizon radius $r_0$
 \be\label{X}
 X|_{r=r_0}=0, \qquad X'|_{r=r_0}=0.
\ee
The equations  (\ref{X}) relate the mass, the rotation parameters, and $r_0$ \cite{kerr-ads}.  Then the coordinates are redefined
\bea\label{limit 2n}
&&
r\rightarrow r_0+\lambda rr_0, \qquad \phi_i \rightarrow \phi_i+\alpha_i \beta t, \qquad \alpha_i=\frac{a_i}{r_0^2+a_i^2},
 \nonumber\\[2pt]
&&
\qquad \qquad t\rightarrow \beta t, \qquad \beta=\frac{\prod_{i=1}^{n-1}(r_0^2+a_i^2)}{\lambda r_0 V},
\eea
where $V=\frac{1}{2}X''|_{r=r_0}$, and the limit $\lambda \rightarrow 0 $ is taken which yields \cite{kerr-ads}
 \bea\label{nearhormetric2n}
 &&
 ds^2=\frac{\widetilde{U}}{V} \Big(-r^2 dt^2+\frac{dr^2}{r^2} \Big)+\sum_{\alpha=1}^{n-1}\frac{\widetilde{U}_\alpha}{X_\alpha}dx_\alpha^2
  \nonumber\\[2pt]
&&
\qquad \qquad +\sum_{\alpha=1}^{n-1} \frac{X_\alpha}{\widetilde{U}_\alpha} \Big[\frac{2r_0 r\prod_{\beta}(r_0^2+x_\beta^2)}{V(r_0^2+x_\alpha^2)} dt+\sum_{i=1}^{n-1}\frac{(r_0^2+a_i^2) \gamma_i}{(a_i^2-x_\alpha^2)\varepsilon_i} d\phi_i \Big]^2.
\eea
Here we denoted $\widetilde{U}=U|_{r=r_0}$, $\widetilde{U}_\alpha=U_\alpha|_{r=r_0}$. It is straightforward to verify that Eq. (\ref{nearhormetric2n}) is a vacuum solution of the Einstein equations.

Near the horizon the isometry group is enhanced.  In addition to time translation and rotations
  \be\label{trans}
 t'=t+\epsilon, \qquad \phi_i'=\phi_i+\epsilon_i,
 \ee
it includes the dilatation
 \be\label{dil}
 t'=t+t \delta, \qquad r'=r-r \delta
\ee
and the special conformal transformation
 \be\label{special 2n}
 t'=t+\Big(\frac{1}{r^2} + t^2 \Big) V\sigma, \qquad r'=r-2 r t V \sigma, \qquad \phi'_i=\phi_i-\frac{c_i}{r}\sigma,
\ee
where $i=1,\dots,n-1$ and the constants  $c_i$  read
\be\label{conform koeff}
c_i=\dfrac{4 r_0a_i \prod_{j=1}^{n-1}(r_0^2+a_j^2)}{(r_0^2+a_i^2)^2}.
 \ee

 The  Killing tensors, which characterize the near-horizon geometry (\ref{nearhormetric2n}), are derived from (\ref{killing}) by taking into account the relation between the metrics (\ref{metr}) and (\ref{me2n}), by applying the coordinate redefinition (\ref{limit 2n}), and by taking the limit $\lambda\rightarrow0$
\bea \label{diverse killing}
&&
\widetilde{K}_{(k)}=\sum_{\mu=1}^{n-1}\frac{\widetilde{A}_{\mu}^{(k)} (x_\mu^2+r_0^2)^2}{X_\mu \widetilde{U}_\mu} \Bigg[\sum_{i=1}^{n-1} \frac{a_i \prod_{j=1}^{n-1}(x_\mu^2-a^2_j)}{(x_\mu^2-a_i^2)(r_0^2+a_i^2)}  \partial_{\phi_i} \Bigg]^2+\sum_{\mu=1}^{n-1}\frac{\widetilde{A}_\mu^{(k)} X_\mu}{\widetilde{U}_\mu} (\partial_{x_\mu})^2
\nonumber\\[2pt]
&& \qquad \qquad - \frac{\widetilde{A}_n^{(k)}}{V\widetilde{U} r^2} \Bigg[ V \partial_t-2rr_0 \sum_{i=1}^{n-1}\frac{a_i \prod_{j=1}^{n-1}(r_0^2+a_j^2)}{(r_0^2+a_i^2)^2} \partial_{\phi_i} \Bigg]^2+\frac{A_n^{(k)} r^2 V}{\widetilde{U}} (\partial_r)^2.
\eea
Here $\widetilde{A}_\mu^{(k)}$ are the functions   $A_\mu^{(k)}$  given in (\ref{funct 2n}) with $x_n=\texttt{i}r_0$ and $k=0,\dots,n-1$.

In order to demonstrate that one of the near-horizon Killing tensors is reducible, let us consider the following linear combination:
 \be\label{comb}
 Q^{AB}=\sum_{k=0}^{n-1} r_0^{2(n-1-k)}K^{AB}_{(k)},
 \ee
where $A,B=1, \dots, 2n$.
It is straightforward to verify that $Q^{AB}$ can be decomposed into the Killing vectors
 \bea\label{privod}
 &&
Q^{AB}=-\frac{1}{2}(k_{(1)}^{A} k_{(3)}^{B}+k_{(3)}^{A} k_{(1)}^{B})+\frac{1}{V}k_{(2)}^A k_{(2)}^B
 \nonumber\\[2pt]
&&
\qquad -\frac{1}{4V}\Big(\sum_{l=4}^{n+2}c_{l-3}k_{(l)}^A\Big) \Big(\sum_{l=4}^{n+2}c_{l-3}k_{(l)}^B\Big),
\eea
where $k_{(1)}^A$ denote components of the Killing vector corresponding to the time translations, $k_{(2)}^A$ are linked to the dilatations, $k_{(3)}^A$ are associated with the special conformal transformations and the remaining ones are related to the shifts of the azimuthal angular variables $\phi_i$. The constants $c_j$, which enter the right hand side of (\ref{privod}), are taken from (\ref{conform koeff}). Thus one Killing tensor can be expressed in terms of the others and a specific quadratic combination of the Killing vectors.\footnote{That there exists only one combination like (\ref{privod}) can be seen as follows. First it is verified that the tensors
(\ref{diverse killing}) are linearly independent. So are the vectors (\ref{trans})--(\ref{special 2n}).
Because the $(x_\alpha)$--components of the vectors with $\alpha=1,\dots, n-1$ are equal to zero, the corresponding components of quadratic combinations formed out of them vanish as well.
In order to reduce a combination of the Killing tensors to a quadratic combination of the Killing vectors, one has to ensure that
its $(x_\alpha,x_\beta)$--components with $\alpha,\beta=1, \dots, n-1$ vanish. It is straightforward to verify that this condition is equivalent to solving a system of linear algebraic equations whose general solution is given above in (\ref{comb}).} Similar results for $d=4$ have been obtained earlier in
\cite{g2,go,jr,fr}.

The fact that one of the Killing tensors turns out to be reducible near the horizon does not imply that the particle model on such a background fails to be completely integrable. As was mentioned above, near the horizon the isometry group is enhanced to include the dilatations and the special conformal transformations. Thus, as far as the issue of the complete integrability is concerned, one of the Killing tensors is lost while two extra Killing vectors appear. So the model is in fact minimally superintegrable. Similar situation occurs for the case of $d=2n+1$ which we discuss in the next section.

\vspace{0.5cm}

\noindent
{\bf 3. Near-horizon Killing tensors in $d=2n+1$}

\vspace{0.3cm}

Rewritten in the coordinates introduced in \cite{general kerr}, the rotating black hole solution of the Einstein equations in $d=2n+1$ reads
 \bea\label{mmetr2n+1}
&&
ds^2=\frac{U}{X} dr^2+\sum_{\a=1}^{n-1}\frac{U_\a}{X_\a} dx_\a ^2-\frac{X}{U}
\Big [dt-\sum_{i=1}^{n-1}a_i^2 \gamma_i\frac{d\phi_i}{\varepsilon_i}\Big ]^2
 \nonumber\\[2pt]
&&
\qquad \qquad +\sum_{\a=1}^{n-1} \frac{X_\a}{U_\a}\Big[dt-\sum_{i=1}^{n} \frac{a_i^2 \gamma_i(r^2+a_i^2)}{a_i^2-x_\a^2} \frac{d\phi_i}{\varepsilon_i}\Big]^2+
\frac{\prod_{k=1}^{n} a_k^2}{r^2 \prod_{\alpha=1}^{n-1}x_\alpha^2} \Big[dt-\sum_{i=1}^{n} (r^2+a_i^2) \gamma_i \frac{d \phi_i}{\varepsilon_i} \Big]^2,
 \nonumber\\[2pt]
&&
 U=\prod_{\a=1}^{n-1} (r^2+x_\a^2),   \qquad U_\a=-(r^2+x_\a^2) \prod_{\begin{subarray}{c} \beta=1\\\a\ne \beta\end{subarray} }^{n-1} (x_\beta^2-x_\a^2),
  \nonumber\\[2pt]
&&
 \gamma_i=\prod_{\a=1}^{n-1} (a_i^2-x_\a^2), \qquad \varepsilon_i=a_i \prod^{n}_{\begin{subarray}{c} k=1\\k\ne i\end{subarray} } (a_i^2-a_k^2),
   \nonumber\\[2pt]
&&
 X_\a=\frac{1}{x_\alpha^2}\prod_{k=1}^{n} (a_k^2-x_\a^2),\qquad  X=\frac{1}{r^2}\prod_{k=1}^{n} (r^2+a_k^2)-2M,
\eea
where $t$ is the temporal coordinate, $r$ is the radial coordinate and $x_\alpha$ with $\alpha=1,\dots,n-1$ and $\phi_i$ with $i=1,\dots,n$ are related to the angular variables.
$M$ stands for the mass and $a_i$ denotes the rotation parameters.
As in the even-dimensional case, the coordinate system will be used which is obtained by the linear change of the coordinates
 \bea\label{linear redefinition 2n+1}
&&
B_{j}^{(k)}\equiv\sum_{\begin{subarray}{c} l_1<l_2\dots<l_k \\j\ne {l_1,\dots,l_k}\end{subarray} } a_{l_1}^2 a_{l_2}^2 \dots a_{l_k}^2, \qquad \frac{\phi_j}{a_j}=\sum_{k=0}^{n-1} B{_j}^{(k)} \psi_{k+1},
\nonumber\\[2pt]
&&
B^{(k)}\equiv\sum_{l_1<l_2\dots<l_k} a_{l_1}^2 a_{l_2}^2 \dots a_{l_k}^2, \qquad t=\psi_0 +\sum_{k=1}^{n}B^{(k)} \psi_k,
\eea
where $x_n=\texttt{i}r$. In these coordinates the metric (\ref{mmetr2n+1}) reads \cite{general kerr}
 \bea\label{metrika 2n+1}
 &&
 \qquad \qquad  ds^2=\sum_{\mu=1}^{n} \ \bigg\{ \frac{dx_\mu}{Q_\mu}+Q_\mu \Big(\sum_{k=0}^{n-1} A_{\mu}^{(k)} d\psi_k \Big)^2 \bigg\} - \frac{c}{A^{(n)}} \Big(\sum_{k=1}^{n}A^{(k)} \partial_{\psi_k} \Big)^2,
\nonumber\\[2pt]
&&
Q_\mu=\frac{X_\mu}{U_\mu}, \qquad A_{\mu}^{(k)}=\sum_{\begin{subarray}{c} \nu_1<\nu_2\dots<\nu_k \\\mu\ne \nu\end{subarray} }^{n} x_{\nu_1}^2 x_{\nu_2}^2 \dots x_{\nu_k}^2, \qquad A^{(k)}=\sum_{\nu_1<\nu_2\dots<\nu_k }^{n} x_{\nu_1}^2 x_{\nu_2}^2 \dots x_{\nu_k}^2
\nonumber\\[2pt]
&&
U_\mu=\prod_{\begin{subarray}{c} \nu=1\\\mu\ne \nu\end{subarray} }^{n} (x_\nu^2-x_\mu^2), \qquad X_\mu =\frac{1}{x_\mu^2}\prod_{k=1}^{n} (a_k^2-x_\mu^2)+2M \delta_{\mu n},
\qquad c=\prod^{n}_{k=1}a_k^2,
\eea
with $A_{\mu}^{(0)}=A^{(0)}\equiv1$.

As in the preceding case, there are $n$ functionally independent second rank Killing tensors \cite{desert}
\be\label{killing 2n+1}
K_{(k)}=\sum_{\mu=1}^{n}\Big[\frac{A_{\mu}^{(k)}}{X_\mu U_\mu} \Big(\sum_{j=0}^{n} (-x^2_{\mu})^{n-1-j} \partial_{\psi_j} \Big)^2+A_{\mu}^{(k)} Q_\mu(\partial_{x_\mu})^2\Big]-\frac{A^{(k)}}{c A^{(n)}}(\partial_{\psi_n})^2,
\ee
where $k = 0, \dots, n - 1$ and $K_{(0)}$ denotes the metric inverse to (\ref{metrika 2n+1}).

In order to construct the near-horizon geometry, the extremal case is considered
 \be
 X|_{r=r_0}=0, \qquad X'|_{r=r_0}=0,
\ee
where $r_0$  is the horizon radius. Then the coordinate redefinition is used
\bea\label{limit 2n+1}
&&
r\rightarrow r_0+\lambda rr_0, \qquad \phi_i \rightarrow \phi_i+\alpha_i \beta t, \qquad \alpha_i=\frac{a_i}{r_0^2+a_i^2},
 \nonumber\\[2pt]
&&
\qquad \qquad t\rightarrow \beta t, \qquad \beta=\frac{\prod_{i=1}^{n}(r_0^2+a_i^2)}{\lambda r_0^3 V}
\eea
and the limit $\lambda\rightarrow 0$ is taken. These yield \cite{kerr-ads}
\bea\label{nearhormetric2n+1}
 &&
 ds^2=\frac{\widetilde{U}}{V} \Big(-r^2 dt^2+\frac{dr^2}{r^2} \Big)+\sum_{\alpha=1}^{n-1}\frac{\widetilde{U}_\alpha}{X_\alpha}dx_\alpha^2
  \nonumber\\[2pt]
&&
 \qquad +\sum_{\alpha=1}^{n-1} \frac{X_\alpha}{\widetilde{U}_\alpha} \Big[\frac{2r_0 r\prod_{\beta}(r_0^2+x_\beta^2)}{V(r_0^2+x_\alpha^2)} dt+\sum_{i=1}^{n}\frac{a_i^2(r_0^2+a_i^2) \gamma_i}{(a_i^2-x_\alpha^2)\varepsilon_i} d\phi_i \Big]^2
  \nonumber\\[2pt]
&&
\qquad +\frac{\prod_{k=1}^{n} a_k^2}{r_0^2 \prod_{\alpha=1}^{n-1}x_\alpha^2} \Big[\frac{2 r \prod_{\beta}^{n-1}(r_0^2+x_\beta^2)}{r_0 V} dt+\sum_{i=1}^{n} (r_0^2+a_i^2) \gamma_i \frac{d\phi_i}{\varepsilon_i}\Big]^2,
 \eea
where $V=\frac{1}{2}X''|_{r=r_0}$.  As in the preceding section, the tilde over a function implies that it is evaluated at $r=r_0$.

For $d=2n+1$, the near horizon conformal symmetry transformations are realized as in Eqs. (\ref{trans})-(\ref{special 2n}) above with the only difference that now the coefficients $c_i$ have the form
\be\label{conform koeff 2n+1}
c_i=\dfrac{4 a_i \prod_{j=1}^{n-1}(r_0^2+a_j^2)}{r_0 (r_0^2+a_i^2)^2},
\ee
with $i=1, \dots, n$.

The second rank Killing tensors which characterize the near-horizon geometry are derived from (\ref{killing 2n+1}) in three steps. First, the redefinition (\ref{limit 2n+1}) is used.
Then the link between (\ref{mmetr2n+1}) and (\ref{metrika 2n+1}) is taken into account. Finally, the limit $\lambda\rightarrow0$ is taken. The result reads
\bea \label{diverse killing 2n+1}
&&
\widetilde{K}_{(k)}=\sum_{\mu=1}^{n-1}\frac{\widetilde{A}_{\mu}^{(k)}(x_\mu^2+r_0^2)^2}{x_\mu^4X_\mu \widetilde{U}_\mu}  \Bigg[\sum_{i=1}^{n} \frac{a_i \prod_{j=1}^{n}(x_\mu^2-a^2_j)}{(x_\mu^2-a_i^2)(r_0^2+a_i^2)}  \partial_{\phi_i} \Bigg]^2+\sum_{\mu=1}^{n-1}\frac{\widetilde{A}_\mu^{(k)} X_\mu}{\widetilde{U}_\mu} (\partial_{x_\mu})^2
\nonumber
\eea
\bea
&& \qquad - \frac{\widetilde{A}_n^{(k)}}{V\widetilde{U}_n r^2 r_0^2} \Bigg[ r_0V \partial_t-2r \sum_{i=1}^{n}\frac{a_i \prod_{j=1}^{n}(r_0^2+a_j^2)}{(r_0^2+a_i^2)^2} \partial_{\phi_i} \Bigg]^2
\nonumber\\[2pt]
&&
\qquad \qquad    +\frac{\widetilde{A}_n^{(k)} r^2 V}{\widetilde{U}} (\partial_r)^2-
\frac{\widetilde{A}^{(k)}r_0^4}{\widetilde{A}^{(n)}} \Big [\sum^{n}_{i=1}b_i \partial_{\phi_i}\Big]^2,
\eea
where $k=0,\dots,n-1$ and the functions $\widetilde{A}_\mu^{(k)}$, $\widetilde{A}^{(k)}$ are obtained from  $A_\mu^{(k)}$, $A^{(k)}$ in Eq. (\ref{metrika 2n+1}) by setting $x_n=\texttt{i}r_0$. The constants  $b_i$, which appear in the previous expression, read
\be\label{b_i}
b_i=\frac{\prod^{n}_{j=1}a_j}{a_i(r_0^2+a_i^2)}.
\ee

The fact that one of the near horizon Killing tensors is reducible is demonstrated by first considering the specific linear combination
\be
 Q^{AB}=\sum_{k=0}^{n-1} r_0^{2(n-1-k)}K^{AB}_{(k)},
 \ee
where $A,B=1, \dots, 2n+1$ and then checking that the latter can be decomposed into the Killing vectors
 \bea
 &&
 Q^{AB}=-\frac{1}{2}(k_{(1)}^{A} k_{(3)}^{B}+k_{(3)}^{A} k_{(1)}^{B})+\frac{1}{V}k_{(2)}^A k_{(2)}^B-\frac{1}{4V}\Big(\sum_{l=4}^{n+3}c_{l-3}k_{(l)}^A\Big) \Big(\sum_{l=4}^{n+3}c_{l-3}k_{(l)}^B\Big)
\nonumber\\[2pt]
&&
\qquad \qquad +r_0^2 \Big(\sum_{l=4}^{n+3}b_{l-3} k_{(l)}^A\Big) \Big(\sum_{l=4}^{n+3}b_{l-3} k_{(l)}^B\Big),
\eea
where $c_i$, $b_i$ are given in (\ref{conform koeff 2n+1}), (\ref{b_i}) above and $k_{(1)}^A$, $k_{(2)}^A$, $k_{(3)}^A$ correspond to the time translations, dilatations and special conformal transformations, respectively, while the remaining Killing vectors are related to the shifts of the azimuthal angles $\phi_i$.  As in the preceding section, we thus conclude that among the near horizon second rank Killing tensors only one is reducible, while the remaining ones prove to be functionally independent.

\vskip 0.5cm
\noindent
{\bf Acknowledgements}\\

\noindent
This work was supported by the Dynasty Foundation and by the RFBR grants 14-02-31139-Mol and 13-02-90602-Arm.

\end{document}